\magnification=1200
\pretolerance 2000
\baselineskip=28pt
\catcode`@=11
\def\vfootnote#1{\insert\footins\bgroup\baselineskip=20pt
  \interlinepenalty\interfootnotelinepenalty
  \splittopskip\ht\strutbox 
  \splitmaxdepth\dp\strutbox \floatingpenalty\@MM
  \leftskip\z@skip \rightskip\z@skip \spaceskip\z@skip \xspaceskip\z@skip
  \textindent{#1}\footstrut\futurelet\next\fo@t}
\skip\footins 20pt plus4pt minus4pt
\def\footstrut{\vbox to2\splittopskip{}}
\catcode`@=12 
\def\folio{\ifnum\pageno=0\else\ifnum\pageno<0 \romannumeral-\pageno
\else\number\pageno \fi\fi}
\def\buildurel#1\under#2{\mathrel{\mathop{\kern0pt #2}\limits_{#1}}}
\vglue 24pt plus 12pt minus 12pt
\input amssym.def
\input amssym.tex

\centerline {{\bf Ionization of a Model Atom by Perturbations of
the Potential}\footnote*{Research Supported by AFOSR Grant F49620-98-1-0207}}
\bigskip
\centerline {by}
\centerline {Alexander Rokhlenko and Joel L. Lebowitz}
\centerline {Department of Mathematics and Physics}
\centerline {Hill Center, Busch Campus}
\centerline {Rutgers University}
\centerline {New Brunswick, NJ 08903}
\bigskip
\bigskip\bf
\centerline {ABSTRACT}
\medskip\rm
We study the time evolution of the wave function of a particle bound by 
an attractive $\delta$-function potential when
it is subjected to time dependent variations of the binding strength
(parametric excitation). The
simplicity of this model permits certain nonperturbative calculations to be
carried out analytically both in one and three dimensions. Thus 
the survival probability of bound state $|\theta(t)|^2$, following a pulse
of strength $r$ and duration $t$, behaves as 
$|\theta(t)|^2 -|\theta(\infty)|^2 \sim t^{-\alpha}$, with both
$\theta(\infty)$ and  $\alpha$
depending on $r$. On the other hand a sequence of strong short pulses
produces an exponential decay over an intermediate time scale. 
\bigskip
\bigskip{\bf1. Introduction}

While there has been much progress in our understanding of the
processes leading to the ionization of atoms and/or the dissociation
of molecules subjected to time dependent fields, the mathematical
difficulties presented are such that there are no explicitely solvable
models for transitions form a bound into the continuum [1--15]. This
motivates us to investigate here the ionization probability of a
particle bound by an attractive point $\delta$-function potential in
one dimension [5,6,15] and a spherically smeared out $\delta$-function
in three dimensions.  We obtain explicit expressions for the
ionization probability and for the energy distribution of the ejected
electrons for certain time dependent parametric excitations, i.e.
when we suddenly change the value of the coupling constant for a time
interval $t$.  {}For such changes the survival probability of the bound
state shows no regime of exponential decay but approaches its
asymptotic value as a power law [11--13].  The situation is different
for periodic forcing with short pulses which is also treated here and
more generally in [16]. The survival probabilities now include
intermediate exponential regimes followed by power law asymptotics.
\bigskip{\bf General formulation}

We consider first the one-dimensional system with unperturbed 
Hamiltonian [5,6,15]$$ 
H_0=-{\hbar^2\over 2m}{d^2\over dx^2}-g\delta (x),\ g>0,\ \
-\infty<x<\infty.\eqno(1)$$ 
$H_0$ has a single bound state $$
u_b(p,x)=\sqrt{p}e^{-p|x|},\ \ p={m\over\hbar^2}g\eqno(2)$$ 
with energy
$-E_0=-\hbar\omega_0=-\hbar^2p^2/2m$ and a continuous uniform spectrum on the
positive real line, with generalized eigenfunctions
$$u(p,k,x)={1\over\sqrt{2\pi}}\left (e^{ikx}-{p\over p+i|k|}e^{i|kx|}
\right ), \ \ -\infty<k<\infty,\eqno(3)$$
and energies $\hbar^2k^2/2m$ (with multiplicity two for $k\not =0$).
Here $u_b$ is normalized to 1 and $u(k,x)$ to $\delta (k-k').$

Beginning at some initial time, say $t=0$, a perturbing potential
$V(x,t) = -R(t) \delta(x)$ 
is applied to the system, i.e.\ we change  the parameter $g$ in
$H_0$,
$$ g\to g+R(t),\ \ t\geq 0.\eqno(4)$$
We note here that the matrix
elements, $|\langle u_b| V |k \rangle|^2 = R^2(t) {p \over 2\pi} k^2/(p^2 +
k^2)$ which vanishes as $k\to 0$ and approaches $R^2(t) {p \over 2\pi}$ 
as $|k| \to \infty$.  This implies in
particular that the integral of the transition matrix over all $k$ is infinite.

To solve the time dependent Schr{\"o}dinger equation,
 $$ i\hbar{\partial \psi
(x,t)\over\partial t}=H_0\psi (x,t)-R(t)
\delta (x)\psi(x,t),\ \ t\geq 0\eqno(5)$$
we expand $\psi (x,t)$ for $t\geq 0$
in the complete set of functions $u$:$$
\psi (x,t)=\theta (t)u_b(p,x)e^{i{\hbar p^2\over2m}t}+\int_{-\infty}^
{\infty}\Theta (k,t)u(p,k,x)e^{-i{\hbar k^2\over2m}t}dk,\ \ t\geq 0\eqno(6)$$
and monitor the evolution of $\theta (t)$ and  $\Theta (k,t)$ starting from
the initial bound state $\theta (0) = 1,\ \Theta (k,0) = 0$.

The ionization
probability at time $t$ caused by a pulse, which coincides with
$R(t^\prime)$ for 
$t^\prime<t$ and vanishes for $t^\prime\geq t$, is given by
$$
P(t)=1-|\theta (t)|^2=\int_{-\infty}^{\infty}|\Theta (k,t)|^2dk,\eqno(7)$$
while $|\theta(t)|^2$ is the survival probability.

This model can be extended to a three dimensional shell-like delta function
potential. The Hamiltonian, $$ H_0=-{\hbar^2\over 2m}\Delta-g\delta (r-a),\ \
a>0, \quad r = |{\bf r}|,\ {\bf r} \in {\Bbb R}^3 \eqno(8)$$ 
has bound states with angular momentum $l$ for all
$l=0,1,...$ such that $l<mga/\hbar^2 -1/2$.  The time dependent
perturbation is now of the form $V(r,t) = -R(t) \delta(r-a)$.  The results
for three dimensions, which are similar to those in one dimension are
described in Section 5.  This follows calculations of $\theta(t)$ and
$\Theta(k,t)$ in one dimension.
\vskip1cm
{\bf 2. Integral equation for the 1-d case}

Using the orthonormality of the eigenfunctions (2), (3) and substituting
(6) into (5) yields the following set of equations for the time dependent
amplitudes at $t\geq 0$, $$ i\hbar {d\theta\over dt}=-\sqrt{p}T(t),\ \
\ i\hbar {\partial\Theta\over \partial t}={i|k|\over\sqrt{2\pi}
(p-i|k|)}e^{i{\hbar \over2m}(p^2+k^2)t}T(t),\eqno(9)$$ 
where$$ T(t)=\left [
\sqrt{p}\theta (t)+{1\over \sqrt{2\pi}}\int_{-\infty}^{\infty} {i|k|\over
p+i|k|}e^{-i{\hbar \over2m}(p^2+k^2)t}\Theta (k,t)dk\right ]R(t)
\eqno(10)$$
determines both $\theta$ and $\Theta$: 
$$
\theta (t)=1+i{\sqrt{p}\over \hbar}\int_0^tT(t')dt',\eqno(11)$$
$$\Theta (k,t)={|k|\over \sqrt{2\pi}(p-i|k|)\hbar}\int_0^tT(t') e^{i{\hbar
\over 2m}(p^2+k^2)t'}dt'.\eqno(12)$$ 
Substituting (11) and (12) into (10)
yields an integral equation for $T(t)$ which using dimensionless variables
obtained by setting $\hbar =2m=g/2=1$ (implying $p=1,\ \omega_0=1$) yields$$
\theta (t) =1+2i\int_0^t Y(t') dt',\ \ \Theta (k,t)=\sqrt{2\over \pi}
{|k|\over 1-i|k|}\int_0^t Y(t')e^{i(1+k^2)t'}dt',\eqno(13)$$
where $Y(t)$ is to be found from the integral equation$$ 
Y(t)=\eta (t)\left\{1+\int_0^t [2i+M(t-t')]Y(t') dt'\right \}\eqno(14)$$ 
and $\eta (t)=R(t)/g$. The function $M(s)$ in (14), is given by $$ 
M(s)={2i\over \pi}\int_0^\infty
{u^2e^{-is(1+u^2)}\over 1+u^2}du= {1\over 2}\sqrt{i\over
\pi}\int_s^\infty{e^{-iu}\over u^{3/2}} du.\eqno(15)$$ 
$M(s)$  behaves as $$
M(s)=\cases{{1-i\over \sqrt{8\pi s^3}}e^{-is}+O(s^{-5/2}),\ &when $s\to 
\infty$,\cr
{1+i\over \sqrt{2\pi s}}-i +O(s^{1/2}),\ &when $s\to 0.$}\eqno(16)$$
\bigskip
{\bf 3. Ionization by a rectangular pulse}

We study perturbations having the form of a step function,
$\eta(t)=r$ for $t\geq 0$. The calculation of $P$ at any time $t$ will
then correspond to the ionization probability caused by a pulse of
amplitude $r$ and duration $t$. Substituting the $\eta(x)$ into (14)
and taking the Laplace transform we find, c.f. [16]$$
\tilde Y(s)=\int_0^\infty Y(x)e^{-sx}dx={r\over s-r(i+\sqrt{is-1})},
\eqno(17)$$
where$$
\Im\sqrt{is-1}>0.\eqno(18)$$
Using the inverse transform Eq.(13) has the form$$
\theta (t)=1+{r\over \pi}\int_{-i\infty +\alpha}^{i\infty 
+\alpha }{e^{st}-1\over s[s-r(i+\sqrt{is-1})]}ds,\ \ \alpha >0.\eqno(19)$$

To evaluate (19) we make a cut in the complex plane of $s$ 
along the imaginary axis from $-i
\infty$ to $-i$. In the left half plane bounded by the left
side of the cut and the vertical line from $-i+\alpha$ to $+i\infty +
\alpha$ the integrand in (19) is analytic except for a simple pole at
$s=ir(r+2)$ when $r+1>0$. There are no poles if $r<-1$, i.e. when the 
coefficient 
of the $\delta$-function is positive and the potential for $t > 0$
represents repulsion. The integral along the left half-circle of
infinitely large radius is clearly zero, therefore one may rewrite
(19) as$$
\theta (t)={r\over \pi }\int {e^{st}\over s[s-(i+\sqrt{is-1})]}ds +
2{r+1+|r+1|\over (r+2)^2}e^{ir(r+2)t},$$
with counterclockwise integration around the cut.  
Straightforward manipulations with the integral term allow us to write 
$\theta$ finally in the form$$
\theta (t)={4r^2\over \pi}\int_0^\infty {e^{-i(1+u^2)t}\over (1+u^2)^2
[(r+1)^2+u^2]}u^2du+2{r+1+|r+1|\over (r+2)^2}e^{ir(r+2)t}.\eqno(20)$$
The integral in (20) can be expressed in terms of Fresnel's functions 
and the dependence of the ionization probability on $r$ 
is shown in Figs.1 and 2 where it seen that it is monotone for $r<-1$
but not for $r>1$ so we can have "atomic stabilization" [14,15].

Using (12) and (17) one can calculate $|\Theta(k,t)|^2$ which gives for
$t\geq\tau $ the energy distribution of electrons 
kicked out from the bound state by a pulse of duration $\tau$.
We find in the original units$$
\Theta (k,t)=\sqrt{2p\over \pi}{i|k|(p-q)\over (p-i|k|)}\Bigg \{
{(q+|q|)e^{i{\hbar q^2\over 2m}\tau}\over (q^2+k^2)(p+q)}-{e^{-i{\hbar k^2
\over 2m}\tau}\over (p+i|k|)(q-i|k|)}+$$
$${1\over p+q}\Bigg [{pe^{i{\hbar p^2\over 2m}\tau }\over p^2+k^2}{\rm Erfc}
\left ({i+1\over 2}\sqrt{\hbar p^2\tau\over m}\right )-{|q|e^{i{\hbar q^2
\over 2m}\tau }\over p^2+k^2}{\rm Erfc}\left ({i+1\over 2}\sqrt{\hbar q^2\tau
\over m}\right )\Bigg ]+$$
$${i|k|(p-q)e^{-i{\hbar k^2\over 2m}\tau}\over (p^2+k^2)(q^2+k^2)}
{\rm Erfc}\left ({i-1\over 2}\sqrt{\hbar k^2\tau\over m}\right )\Bigg \},
\quad t \geq \tau\eqno(21)$$
where $q=(1+r)p$ and Erfc$(z)$  denotes the probability integral. 
{}For large $k$,
$|\Theta(k,\tau)|^2$ decays like $k^{-4}$ giving a very long tail to the
energy distribution of the emitted electrons.  In {}Fig.\ 3 we plot
$|\Theta(k,\tau)|^2$ vs. $k$ for several values of $\tau$ when $r=-1$,
i.e.\ when the pulse just destroys the attractive interaction. It is seen
that the longer the pulse the more peaked is the distribution with the
maximum moving towards small values of $k$.

The total energy of the electrons ejected by the pulse is given by 
$$
E(t) = {\hbar^2 \over 2m} \int_{-\infty}^\infty k^2 |\Theta(k,\tau)|^2 dk,
\quad t \geq \tau.$$
{}For measurements made outside the range of the potential
this energy will be the same as the kinetic energy of the emitted
electrons. An analytical evaluation of $E(t)$ 
yields a very long and not very illuminating formula. Instead we present in
{}Fig.\ 4 a numerical plot of $E(t )$ for $r=-1$.  When the pulse
length $\tau \to \infty$, $E(t)$ approaches the value $$ E(\infty)=E_0
\left ({r\over |r+1|+1}\right )^2\left [1+2|r+1|+
2(r+1+|r+1|){(r+1)(r+3)\over (r+2)^2}\right ],$$ which increases linearly
with $|r|$ when $|r|\to\infty $:
$$E(\infty)\to 2E_0|2r+|r||.$$
Attractive long pulses, $r>0$, thus give three times as much energy to the 
ejected electrons than do the repulsive ones, $r<0$.  This is shown in 
{}Fig.\ 5 where it is seen that $E(\infty )$ is monotone for both positive and 
negative $r$.

We note that a rectangular pulse perturbation is a special case
of a sudden jump from the initial Hamiltonian $H_0$ to a new time-independent
Hamiltonian $H_1$ which in return jumps to $H_0$ when the perturbation
ends. The amplitudes $\theta$, and $\Theta$ can thus also be calculated by 
projecting (twice) a new state onto the
old one which is just the evaluation of overlap integrals. The Laplace
method, which gives $\theta (t)$ for general tine dependence of $R(t)$ in
the form
$$
\theta (t)=1+ {1\over \pi}\int_{-i\infty +\alpha}^{i\infty 
+\alpha }{e^{st}-1\over s}\tilde Y(s)ds,\eqno(22) $$
shows that $|\theta (t)|^2$ will have an exponential decay for 
$t\to\infty$ only when $\tilde Y(s)$ has poles in the 
left half of the complex plane $s$. 
\vskip1cm
{\bf Power law decay }

When the pulse length $t$ goes to infinity the integral in (20) vanishes
and in the limit we have$$
|\theta (\infty)|^2=\cases {16(r+1)^2/(r+2)^4, &if $r\geq -1$,\cr
0, &if $r<-1$}.\eqno(23)$$
It is seen from (23) that any two very long pulses  
(at least one of them must be repulsive) 
produce the same ionization if their amplitudes $r$ and $r'$ satisfy the
relation  ${1\over r}+{1\over r'}=-1.$

{}For large $t$ the asymptotics of the integral term
in (20) can be easily found.  Using contour integration we can 
rewrite the integral as$$ 
{1\over (r+1)^2t\sqrt{it}}\int_0^\infty{y^2e^{-y^2}\over(1-iy^2/t)^2
[1-iy^2/t(1+r)^2]}dy$$ 
and integrate by expanding the integrand in powers of $y^2/t$.
Let us first study the case $r=-1$, which corresponds to the perturbation
removing the potential and making the electron evolve for $t > 0$ like a
free particle. 
The decay of the bound state in this case is rather slow:$$ |\theta(t)|^2 =
{4\over \pi t}+O(t^{-2}),\ r=-1.\eqno(24)$$

When both $t$ and $t|r+1|$ are large we get$$
\theta(t) = 2{r+1+|r+1|\over (r+2)^2}e^{itr(r+2)}+{r^2\over 
(r+1)^2t\sqrt{i\pi t}}e^{-it}+O(t^{-5/2}).\eqno(25)$$ 
{}For the survival
probability of the bound state we have$$ 
|\theta (t)|^2\approx
\cases{|\theta (\infty)|^2+{8r^2\cos {[(r+1)^2t]}\over (r+1)(r+2)^2t\sqrt{\pi
t}},&if $r>-1$,\cr {r^4\over (r+1)^4\pi t^3},&if $r<-1$.\cr}\eqno(26)$$ 
Thus for $r \leq -1$, when the evolution takes place with a repulsive
$\delta$-function, the approach to zero of $|\theta(t)|^2$ is like
$t^{-3}$, compared to the $t^{-1}$ decay given in (24) for the free
evolution, see {}Fig.2. Note that the coefficient of $t^{-3}$ becomes 
independent of
$r$ for $|r| >> 1$.  {}For $r> -1$ the approach of $|\theta(t)|^2$ to its
nonvanishing asymptotic value is
oscillatory with an envelope which decays like $t^{-3/2}$.  These
oscillations are very rapid for large $r$ ({}Fig.1), but their amplitude 
is small, of
order $1/r$.  These asymptotic power law decays are in agreement with
general results for the decay of initially localized states, c.f.\
[5--8].
\bigskip
{\bf 4. Ionization by periodic short pulses}

The behavior of $P(t)$ for short pulses of duration $t <<1$, is very 
different for cases when $a=r\sqrt{t}$ is large or small comparing with 1.
Writing $P(t)=P(t,a)$ we analyze Eq.(20) and have in the case of a single 
pulse$$
P(t,a)=4\sqrt{2t\over\pi}\cases{2a^2/3 & for $a<<1$,\cr
1 &for $a>>1$.}\eqno(27)$$

We turn now to the survival probability when we bombard our system with a
whole train of short pulses of duration $\tau<<1$  repeated periodically 
with period $\sigma \sim 1$. Using
(13) yields $$
\theta (n\sigma +\tau)=\theta_n=1+2i\sum_{k=0}^nJ_k^0,\ \ \theta_n=1
\ {\rm if}\ n<0,\eqno(28)$$
where we have defined$$
J_n^m=\int_{n\sigma}^{n\sigma +\tau}(n\sigma +\tau -x)^mY(x)dx.\eqno(29)$$
By integrating Eq.(13) for $\theta(t)$ in $t$ from $t=n\sigma$ to 
$t=n\sigma +\tau$ and using (16) we obtain$$\eqalign{
r^{-1}J_n^m ={\tau^{m+1}\over m+1}\left [1+\sum_{k=0}^{n-1}(2i+
M_{n-k})J_k^0\right ]+\sqrt{i\over \pi}k_mJ_n^{m+1/2} + \cr
{i\over m+1}J_n^{m+1} +{3k_m\over\sqrt{i\pi}(2m+3)}J_n^{m+3/2} +O(
\tau^{m+2}),}\eqno(30)$$
where $M_{n}=M(n\sigma ),\ \ k_m=\sqrt{\pi}\Gamma (m+1)/
\Gamma (m+{3\over 2}).$
The inequality $|\theta_n|\leq 1$ implies $|J_k^0|\leq 1$ and
therefore by integration by parts we get$$
|J_k^m|\leq\tau^m.\eqno(31)$$
Let us eliminate in (30) the term $J_n^{m+1/2}$ by using (30) with 
$m\to m+1/2$ which gives$$
r^{-1}J_n^m =\left ({\tau^{m+1}\over m+1}+2k_mr\sqrt{i\over \pi}{\tau^{m+3/2}
\over 2m+3}\right )\left [1+\sum_{k=0}^{n-1}(2i+M_{n-k})J_k^0\right ]+$$
$$i\left ({1\over m+1}+ {rk_mk_{m+1/2}\over \pi}\right )J_n^{m+1}+
O(\tau^{m+2}).\eqno(32)$$ 
Combining (31) with the estimate $|M(s)|<
\sqrt{\pi /2s^3}$ we have an upper bound on the sum in (32) in the
form $|\sum_{k=0}^{n-1}M_{n-k}J_k^0|<2.4\sigma^{-3/2}\max_{j\in [0,n-1]}
|J_j^0|$. Treating the amplitude $r$ as a quantity of order of unity 
one can immediately improve the upper bounds (31) for $J_n^m$ to:$$
|J_n^m|\sim \tau^{m+1}.\eqno(33)$$
Eq.(33) allows us to rewrite Eq.(32) for $m=0$ as a simple
reccurence$$
J_n^0 =\rho \left (1+2i\sum_{k=0}^{n-1}J_k^0\right )+
\tau^2 f_n(\tau),\eqno(34)$$
where for $\sigma\geq 1,\ r\leq 1$ we have $|f_n(\tau)|<7$ uniformly in 
$n$ and $$
\rho =r\tau\left [1+{4r\sqrt{\tau}(1+i)\over 3\sqrt{2\pi}}\right ].$$ 
Starting with $J_0^0=\rho\sim \tau $ one can find successively    
$J_n^0$ using (34). The terms of such a sequence will be close to the 
corresponding terms of the solution of the simplified equations$$
\tilde J_n=\rho \left (1+2i\sum_{k=0}^{n-1}\tilde J_k\right ),\eqno(35)$$ 
till $|\tilde J_n|>>\tau^2$ and $|\tilde J_n|>>|\rho| n \tau^2$. It is
easy to solve (35) to find $$
\tilde J_n=\rho (1+2i\rho )^n,\eqno(36)$$ 
and therefore $|\tilde J_n|\approx |\rho|e^{-n\gamma }$ where
$\gamma = 8r^2\tau^{3/2}/3\sqrt{2\pi}<<1$. Our condition for approximating 
$J_n^0$ by $\tilde J_n$ has now the form$$
e^{-n\gamma }>>n\tau^2 .\eqno(37)$$
Using $\tilde J_n$ and (28) we find $$
\theta (t)\approx \exp \left ({-\gamma+2ir\tau \over\sigma}t\right ),
\eqno(38)$$
if the duration of train of pulses $t=n\sigma$ is not too long and 
satisfies (37). One can obtain from (37) that the decay of survival 
probality $|\theta (x)|^2$ up to a value $\mu$ is accurately described 
by (38) if $\sqrt{\tau}<2r^2\sqrt{\mu}/\ln{\mu^{-1}}$; 
for $\mu =0.01,\ r\approx 1$ this gives $\sqrt{\tau}<0.04$ and
a train of about 300 pulses. {}For shorter $\tau$ the train can be longer
and the ionization more complete.

{}For longer trains of perturbation the term $\sum_{k=0}^{n-1}M_{n-k}
J_k^0$ in (30) cannot be ignored and it makes the eventual decay
slower with strong oscillations due to interference with the 
eigenfrequency. 
We see that in the exponentially decaying regime the survival probality
is independent of $\sigma$.  This is very different from the case where the
time dependent $\eta(t) = r \sin \omega t$ considered in [16].  In that
case the exponential decay depends strongly on $\omega$.  In our case $\tau
\to 0$ which means that $\eta(t)$ will contain all ranges of frequencies.  
\bigskip
{\bf 5. Three dimensional model}

The Hamiltonian (8) has eigenfunctions in the continuum spectrum$$
\Psi_{l,m} (k,{\bf r})=Y_{l,m}(\theta,\varphi)R_l(k,r),\ \ r\geq 0,\ \ 
0\leq\theta\leq\pi,\ \ 0\leq\varphi\leq 2\pi , $$
where the radial functions are$$
R_l(k,r)=A_l\sqrt{k\over r}J_{l+1/2}(kr)+$$
$$A_l\sqrt{k\over r}\cases{0,\ {\rm if}\ &$r\leq a$,\cr
{\pi i\over 4}QJ_{l+1/2}(ka)[H_{l+1/2}^{(1)}(ka)
H_{l+1/2}^{(2)}(kr)-H_{l+1/2}^{(1)}(kr)H_{l+1/2}^{(2)}(ka)]& $r>a$.\cr}
\eqno(40)$$
The dimensionless parameters $A_l$ normalize $R_l(r)$ to
a $\delta$-function,$$ A_l(k)=\{1+Q\pi
J_{l+1/2}(ka)N_{l+1/2}(ka)+Q^2\pi^2J^2_{l+1/2}(ka)[
J^2_{l+1/2}(ka)+N^2_{l+1/2}(ka)]\}^{-1/2},$$ and the notations for
normalized spherical harmonics $Y_{l,m}$ and Bessel functions are the
usual ones.  The energy corresponding $\psi_{l,m}(k,r)$ is  $\hbar^2k^2/2m$.

The parameter $Q=2mga/\hbar^2$ plays a crucial role for the existence of the
bound states,$$ 
QK_{l+1/2}(p_la)I_{l+1/2}(p_la)=1\eqno(41)$$ 
is the
equation for the energy $-\hbar^2p_l^2/2m$ of all bound $l$-states (they
are of different axial symmetry). The left side of (41) is a monotonically
decreasing function of its argument $\gamma =p_la$ and it is equal to
$Q(2l+1)^{-1}$ when $\gamma =0$, therefore $$ Q> 2l+1$$ is the condition to
have the bound states for all $l'\leq l.$ 

The radial normalized
eigenfunctions can be written in the form $$
R_l^b(r)={B_{l}p_l\over\sqrt{r}}\cases {I_{l+1/2}(p_lr),& if $r\leq a,$\cr
I_{l+1/2}(p_la)K_{l+1/2}(p_lr)/K_{l+1/2}(p_la),& $r > a,$\cr}\eqno(42)$$
where $$ B_l={\sqrt{2}K_{l+1/2}(p_la)\over\sqrt{1-p_laK_{l+1/2}(p_la)[
I_{l-1/2}(p_la)+I_{l+3/2}(p_la)]}}$$ and $I$, $K$ are the modified Bessel
functions.

There are no transitions between states of different angular symmetry if
both the potential in (8) and perturbation $V(t,r)$ are central.  {}For 
simplicity
we consider our three-dimensional model with $Q>1$ in in the s-state.
Dropping the index $l=0$ equation (41) for the energy of the bound state
$-\hbar^2p^2/2m$ 
is$$ Q={2ap\over 1-e^{-2ap}}.$$ The eigenfunctions (42) of the bound
 and the continuum states are
respectively$$
\Psi_b(r)={p^{1/2}\over r\sqrt{\pi(e^{2pa}-1-2pa)}}\cases {\sinh pr,&
if $r\leq a$,\cr e^{-p(r-a)}\sinh pa,& $r > a$,\cr}\eqno(43)$$
$$
\Psi_{0,0} (k,r)={2^{-1/2}\over \pi r\sqrt{1-Q{\sin 2ka\over
ka}+Q^2{\sin^2ka\over k^2a^2}}}\cases {\sin kr,& if $r\leq a$,\cr
\sin kr -Q{\sin ka\over ka}\sin k(r-a),& $r>a$.\cr}\eqno(44)$$

Assuming that the particle is in the bound state $\Psi_b(r)$ at $t=0$
and the perturbation has the form $V(r,t) = -R(t) g\delta(r-a)$, we use 
the method of projections which was described in Section 4 to find the
ionization probability induced by the rectangular pulses $R(t) = rg$ 
for $t> 0$. 
After the end of pulse at $t=\tau$ we have for $\theta (\tau)$ an 
equation similar to (21), $$
\theta (\tau )= {4pq\over (e^{2pa}-1-2pa)(e^{2qa}-1-2qa)}\left [
{e^{(p+q)a}\over p+q}-{pe^{(p-q)a}-qe^{(q-p)a}\over
p^2-q^2}\right ]^2e^{i{\hbar q^2\over 2m}\tau}+$$
$$\eqno(45)$$
$$8p{[(pa-Q_1)\sinh pa+pa\cosh pa]^2\over \pi a^2}\int_0^\infty
{e^{-i{\hbar k^2 \over 2m}\tau}\sin^2 ka\over (p^2+k^2)^2\left ( 1-
Q_1{\sin 2ka\over ka}+Q_1^2{\sin^2 ka\over k^2a^2}\right )}dk,$$ where $q$
is the solution of Eq.(41) with $Q_1=(1+r)Q$ instead of $Q$, ($q$ gives the
energy of new bound state). If $Q_1<1$ the first term in (45) vanishes,
otherwise the square of its absolute value represents the probability
$1-P(\infty)$ of the electron to remain in the bound state when $\tau \to
\infty$. Using the dimensionless time $\omega_0t\to t$ the asymptotics of the
decaying term in (45) when $t\to \infty$ is$$
\theta(t)=\theta (\infty )-\sqrt{2}{[(pa-Q_1)\sinh pa+pa\cosh pa]^2\over
(Q_1-1)^2}{1+i\over t\sqrt{\pi t}}+O(t^{-5/2}),\ \ t\to \infty,\eqno(46)$$
or $$ |\theta (t)|^2\approx |\theta (\infty)|^2+\cases{O(t^{-3/2}),&if
$Q_1>1$,\cr O(t^{-3}),&if $Q_1<1$.\cr}$$ The dimensionality as one can see
changes the character of asymptotics only of the free evolution $(t^{-3/2}\
\ vs\ \ t^{-1/2})$. An interesting case is $Q_1=1$, when the perturbed
Hamiltonian has a ``zero energy bound state''. The
asymptotic behavior of $\theta(t)$ is now given by $$
\theta (t)={4[(pa-Q_1)\sinh pa+pa\cosh pa]^2(1-i)\over
p^2a^2\sqrt{2\pi t}}+O(t^{-3/2}),\ \ t\to \infty\eqno(47)$$
which has the same character as for the free decay in the
one-dimensional model.

In three dimensions the same technique as that used in Section 3 allows us
to derive a one dimensional integral equations similar to (14) for each
pair of quantum numbers $l,\ m\leq l$: $$ T_{l,m}(t)=R(t)a^2\left
[R_l^b(a)\theta_{l,m}(0)+{i\over \hbar}\int_0^t
K_l(t-t')T_{l,m}(t')dt'\right ],\eqno(48)$$ which determines the evolution.
In particular, the amplitude of the bound state develops in time
as$$
\theta_{l,m}(t)=\theta_{l,m}(0)+i{R_l^b(a)\over
\hbar}\int_0^tT_{l,m}(t')dt'.\eqno(49)$$
The function $K_l$ in Eq.(48),$$
K_l(\vartheta)=[R_l^b(a)]^2+\int_0^\infty|R_l(k,a)|^2e^{-i\hbar(k^2
+p_l^2)\vartheta /2m}dk,\eqno(50)$$ is independent of the quantum number
$m$. Each spherical harmonic evolves autonomously and if $\theta_{l,m}$
was zero at $t=0$ it does not change for our perturbation.  
Though the kernel of Eq.(48) even for $l=0$, $$
K_0(\vartheta)={4p\sinh^2pa\over a^2(e^{2pa}-1-2pa)}+{2\over \pi a^2}
\int_0^\infty {e^{-i\hbar {k^2+p^2\over 2m}\vartheta}\sin^2ka\over
1-Q{\sin 2ka\over ka}+Q^2{\sin^2 ka\over k^2a^2}}dk,\eqno(51)$$
cannot be expressed in terms of standard
functions numerical calculations are quite feasible. 
\vskip1cm
{\bf 6. Concluding remarks}

Some general features of our results with possible implications for
realistic systems include:  

a) The ionization probability approaches its asymptotic value as
$t^{-3/2}$ if the electron can be bound in the perturbed state, goes to
zero as $t^{-1}$ if the perturbation makes the electron a free particle, and
as $t^{-3}$ when the perturbation converts the attractive well into a
repulsive one.

b) A finite train of periodically repeated short pulses makes the
survival probability of the bound state decay exponentially without
oscillations.
When the frequency of repetition is comparable with the eigenfrequency 
of the bound state or is lower the decay scales in such a way that only the
total number of pulses is important.

c) The three dimensional potential gives a similar
behavior of the ionization. The free evolution in one dimension corresponds
here to a marginal situation with the "zero-energy" bound state.
\vskip1.5cm
\noindent {\bf Acknowledgments:}  We thank R. Barker, O. Costin,
S. Guerrin, H. Jauslin, A. Soffer and M. Weinstein, for useful discussions.
\vskip1.5cm
\centerline {\bf REFERENCES}

\item{[1]} U. Fano, Nuova Cimenta {\bf 12}, 156 (1935); Phys. Rev. {\bf
124}, 1886 (1961); K. O. Friedricks, Comm. Pure Appl. Math {\bf 1}, 361
(1948); R. H. Dicke, Phys. Rev. {\bf 93}, 99 (1954).

\item{[2]} {\it Atom-Photon Interactions}, by C. Cohen-Tannoudji,
J. Duport-Roc and G. Arynberg, Wiley (1992); {\it Multiphoton Ionization
of Atoms}, S. L. Chin and P. Lambropoulus, editors, Academic Press
(1984).

\item{[3]}  P. M. Koch and K.A.H. van Leeuwen, Physics Reports {\bf
255}, 289 (1995);  R. Bl{\"u}mel and U. Similansky, Z. Phys. D{\bf 6}, 83
(1987); G. Casatti and L. Molinari, Prog. Theor. Phys. (Suppl) {\bf 98},
286 (1989).

\item{[4]} S. Guerin and H.-R. Jauslin, Phys. Rev. A {\bf 55}, 1262 (1997)
and references there; E.V.Volkova, A.M.Popov, and O.V.Tikhonova, Zh.Eksp.
Teor.Fiz. {\bf 113}, 128 (1998)

\item{[5]} Yu. N. Demkov and V. N. Ostrovskii, {\it Zero Range
Potentials and Their Application in Atomic Physics}, Plenum (1988);
S. Albeverio, F. Gesztesy, R. H\o egh-Krohn and H. Holden {\it Solvable
Models in Quantum Mechanics}, Springer-Verlag (1988).  

\item{[6]} . M. Susskind, S. C. Cowley, and E. J. Valeo, Phys.Rev. A
{\bf 42}, 3090 (1994); G. Scharf, K.Sonnenmoser, and W. F. Wreszinski,
Phys.Rev. A {\bf 44}, 3250 (1991); S. Geltman, J. Phys. B: Atom. Molec.
Phys. {\bf 5}, 831 (1977); E. J. Austin, {\it Jour. of Physics} {\bf
B12} 4045 (1979); K. J. LaGattuta, {\it Phys. Rev.} {\bf A40} (1989)
683; A. Sanpera and L. Roso-Franco, {\it Phys. Rev.} {\bf A41} (1990)
6515; R. Robusteli, D. Saladin and G. Scharf, {\it Helv. Phys. Acta. \bf
70} 96 (1997); T. P. Grozdanof, P. S. Kristic and M. H. Mittleman, {\it
Phys. Lett. \bf A149} (1990) 144; J. Mostowski and J. H. Eberly, {\it
Jour. Opt. Soc. Am. \bf B8} 1212 (1991); A. Sanpera, Q. Su and
L. Roso-Franco, {\it Phys. Rev. \bf A47} (1993) 2312.

\item{[7]} H. L. Cycon, R. G. Froese, W. Kirsch and B. Simon
{\it Schr\"odinger Operators} Springer-Verlag  (1987).

\item{[8]} C.-A. Pillet, Comm. Math. Phys. {\bf 102}, 237 (1985) and
{\bf 105}, 259 (1986); K. Yajima, Comm. Math.Phys. {\bf 89}, 331 (1982).

\item{[9]} I. Siegel, Comm. Math. Phys. {\bf 153}, 297 (1993).

\item{[10]} A. Maquet, S.-I. Chu and W. P. Reinhardt, Phys. Rev. A {\bf
27}, 2946 (1983); C. Holt, M. Raymer, and W. P. Reinhardt, Phys. Rev. A
{\bf 27}, 2971 (1983); S.-I. Chu, Adv. Chem. Phys. {\bf 73}, 2799 (1988);
R. M. Potvliege and R. Shakeshaft, Phys. Rev. A {\bf 40}, 3061 (1989).

\item{[11]}  A. Soffer and M. I. Weinstein, Jour. Stat. Phys. {\bf 93},
359--391 (1998).

\item{[12]} G. Garcia-Calder{\'o}n, J. L. Mateos, and M. Moshinsky,
Phys. Rev. Lett. {\bf 74}, 337 (1995); Annals of Physics {\bf 249}, 430--
453 (1996).

\item{[13]} J. Stalker, {\it An Essentially Singular Classical Limit},
preprint, 
Princeton, 1998.

\item{[14]}  A. Fring, V. Kostrykin and R. Schrader, {\it Jour. of
Physics} {\bf B29} (1996 5651; C. Figueira de Morisson Faria, A. Fring
and R. Schrader, {\it Jour. of Physics} {\bf B31} (1998) 449; A. Fring,
V. Kostrykin and R. Schrader, {\it Jour. of Physics} {\bf A30} (1997)
8559.

\item{[15]} C. Figueira de Morisson Faria, A. Fring and R. Schrader, {\it
Analytical treatment of stabilization} preprint physics/9808047 v2.

\item{[16]} O. Costin,  J. L. Lebowitz and A. Rokhlenko, {\it Exact
Results for the Ionization of a Model Quantum System} preprint (1999),
Los Alamos 9905038 and work in preparation.
\vfill\eject
\centerline {\bf FIGURE CAPTIONS}
\vskip1cm
Fig. 1. The survival probability of bound state following the imposition of 
rectangular pulses of duration $t$ for different relative 
amplitudes $r>-1$, see Eqs.(23),(24).
\vskip0.6cm
Fig. 2. The normalized survival probability following the imposition of 
repulsive rectangular pulses, $r<-1$. Plots of $|\theta (t)/\theta_{as} (t)|^2$
vs. $t$, where $|\theta_{as} (t)|^2$ is $4/\pi t$ for $r=-1$
and $(r/r+1)^4/\pi t^3$ if $r<-1$, see Eqs.(24),(26).
\vskip0.6cm
Fig3. Plot of $|\Theta (k,\tau )|^2$ which represents the energy 
distribution of electrons kicked out by the rectangular pulses of
duration $t$, see Eq.(21). 
\vskip0.6cm
Fig 4. The total energy of states of electrons in the continuum spectrum 
ejected by the rectangular pulse of duration $t$ for $r=-1$. 
\vskip0.6cm
Fig. 5. Plot of the electron kinetic energy vs. the amplitude of very long
rectangular pulses.
\vskip0.6cm
\vfill\eject

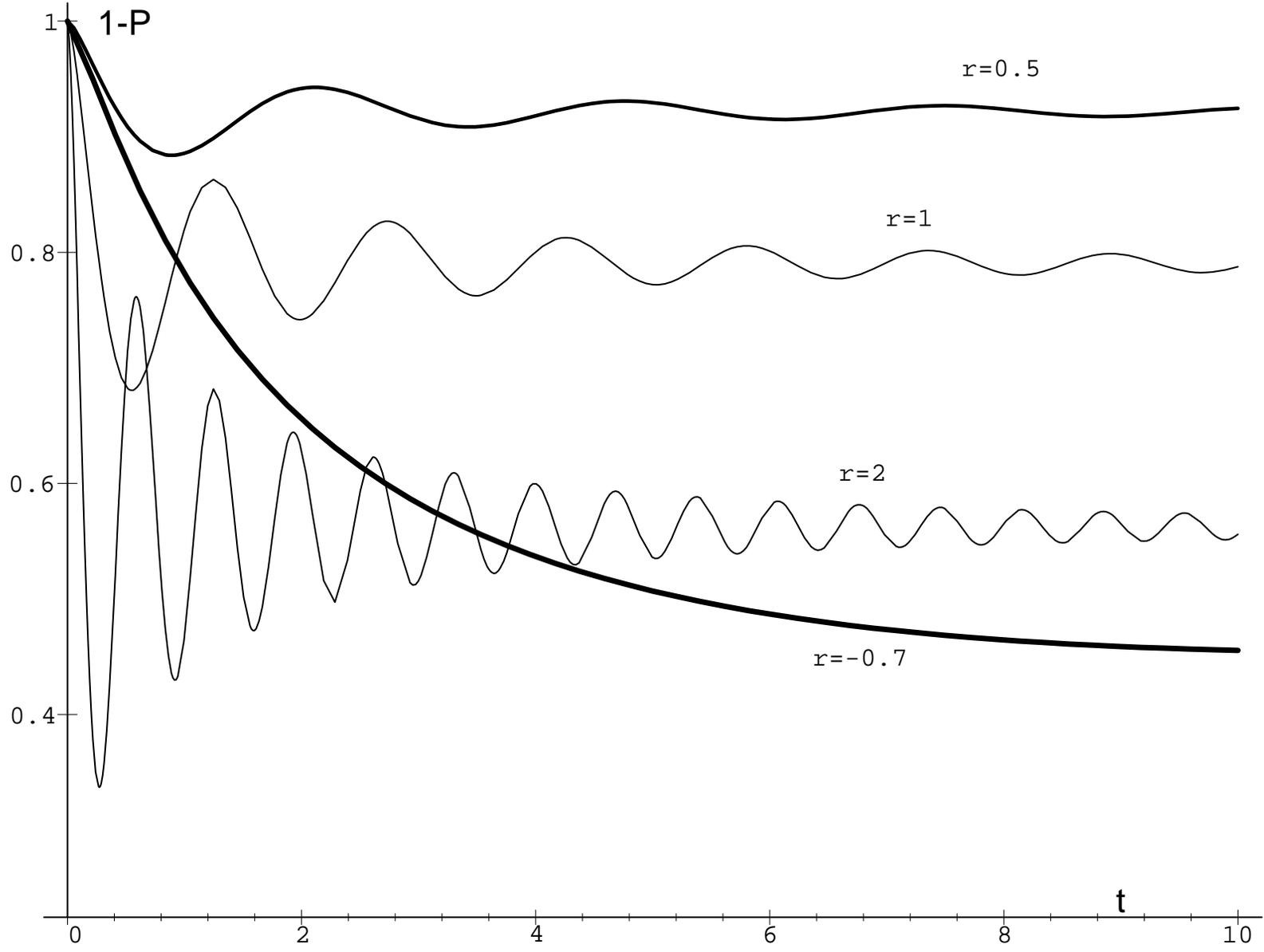
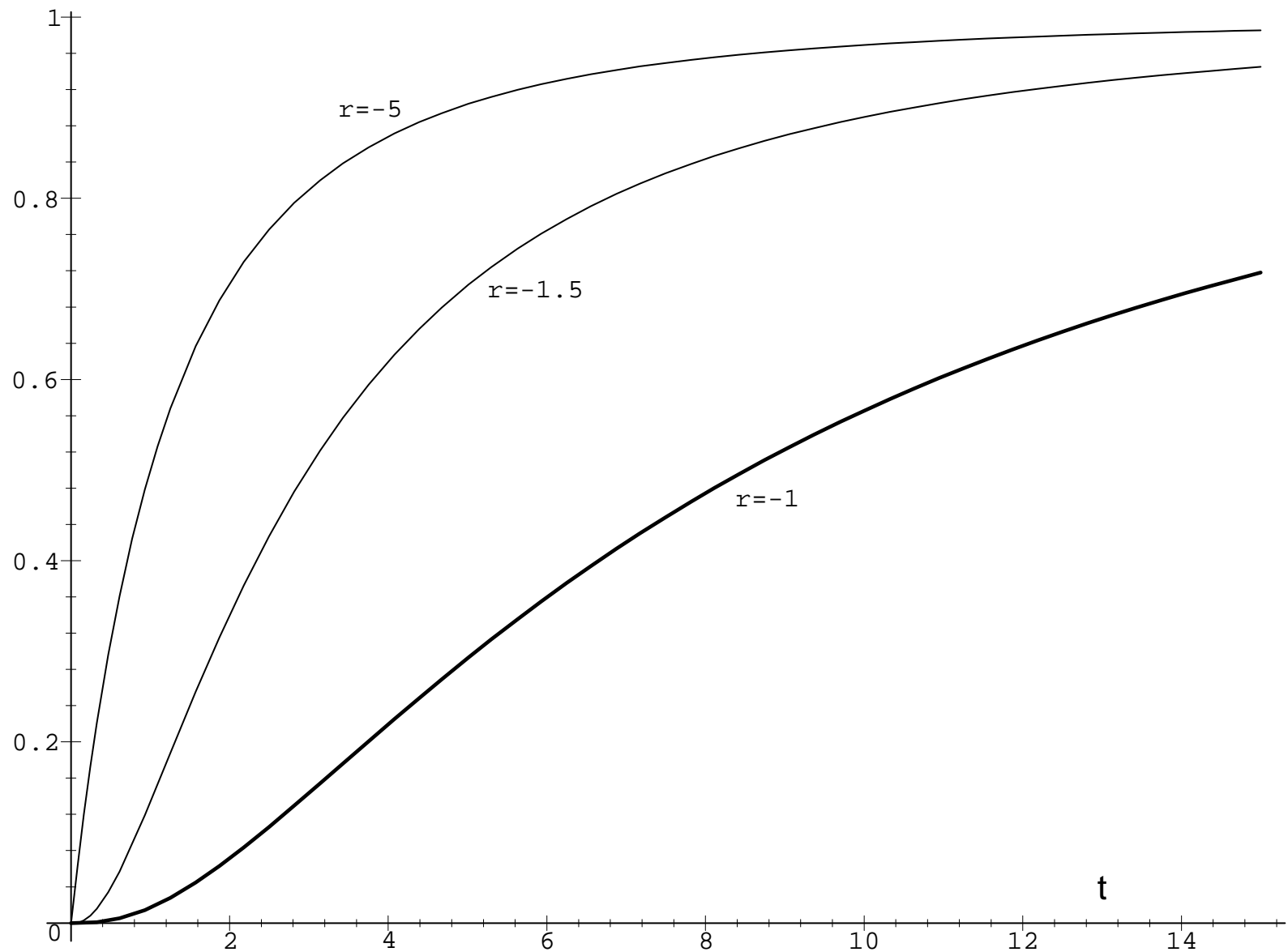
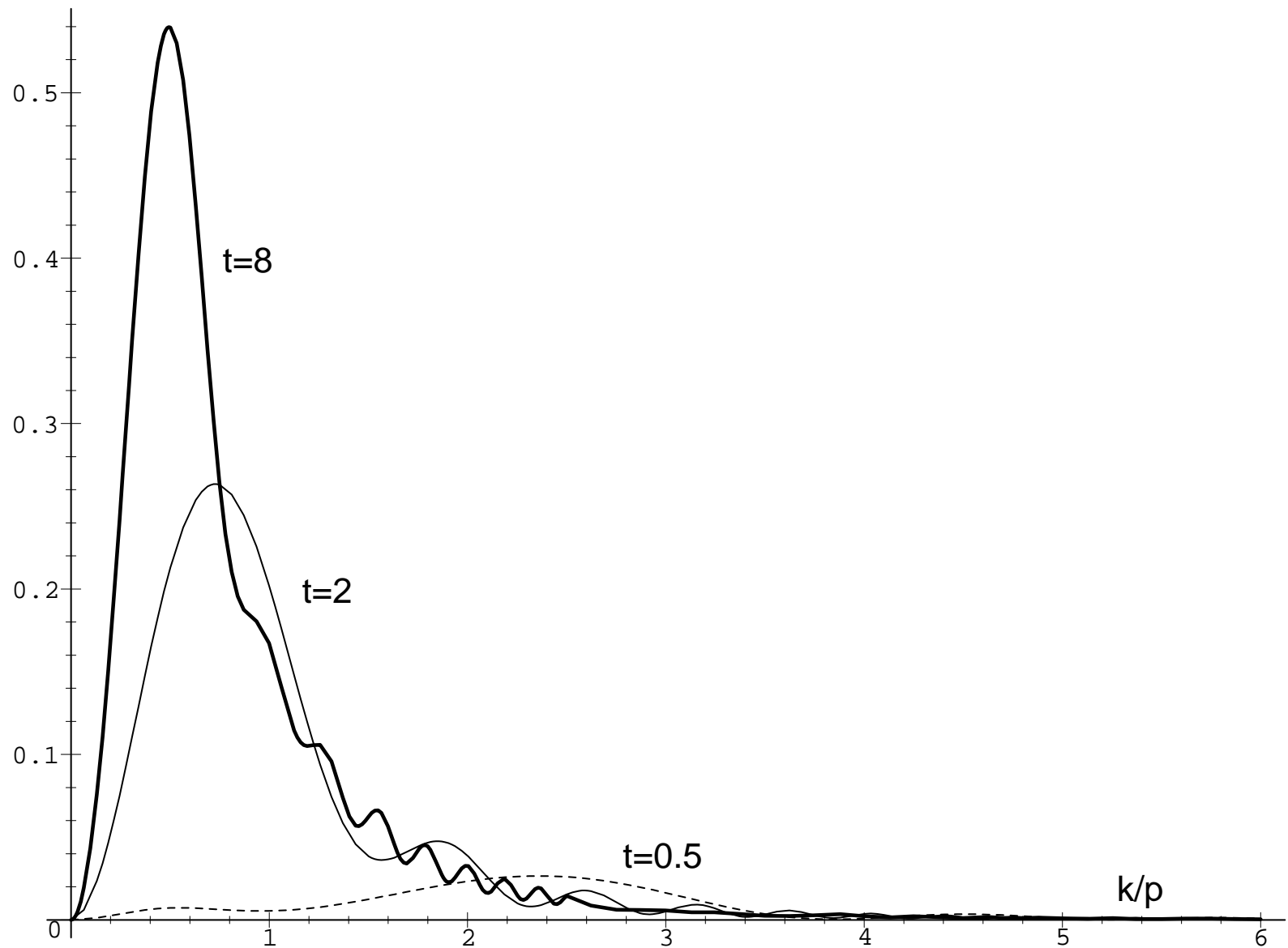
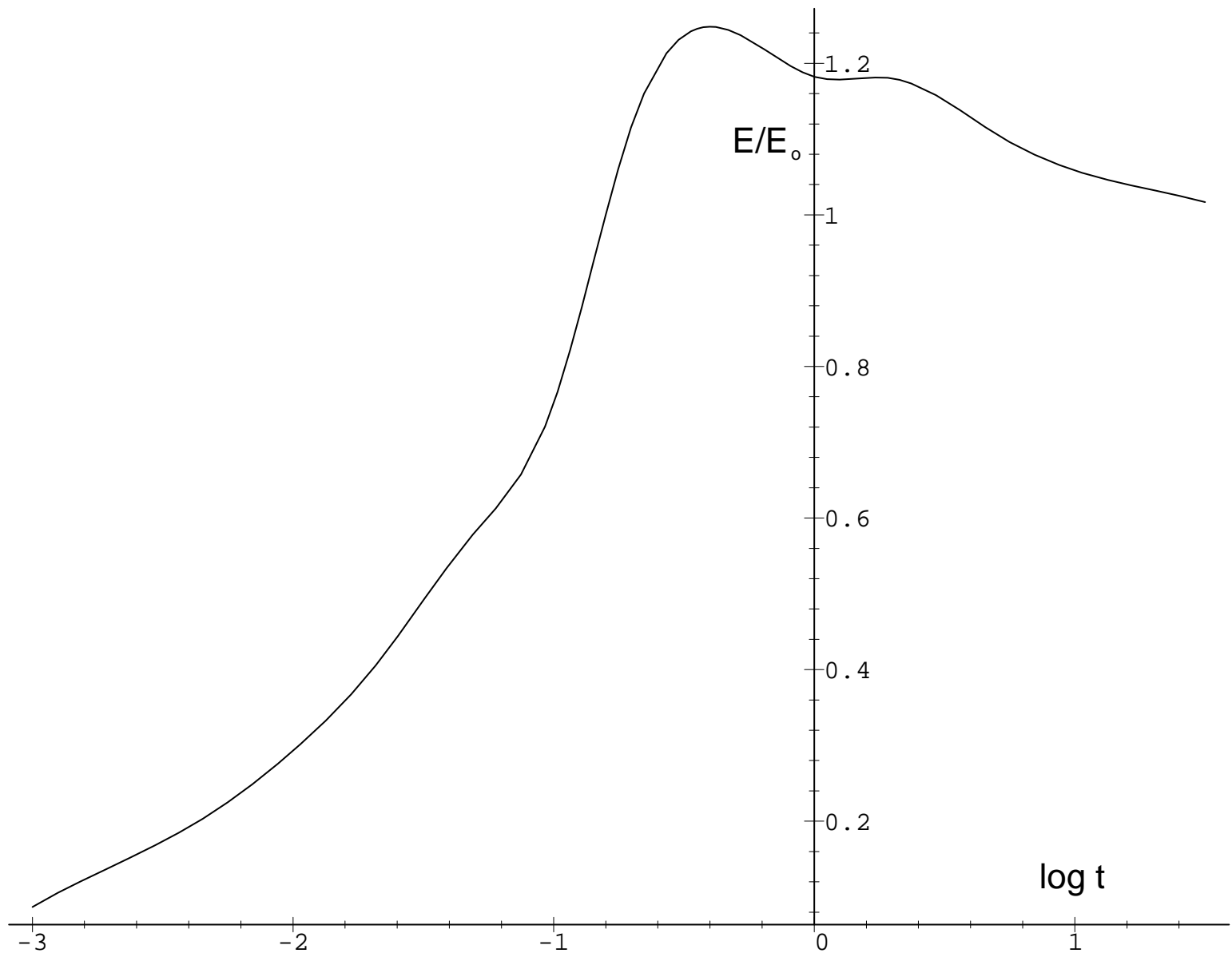
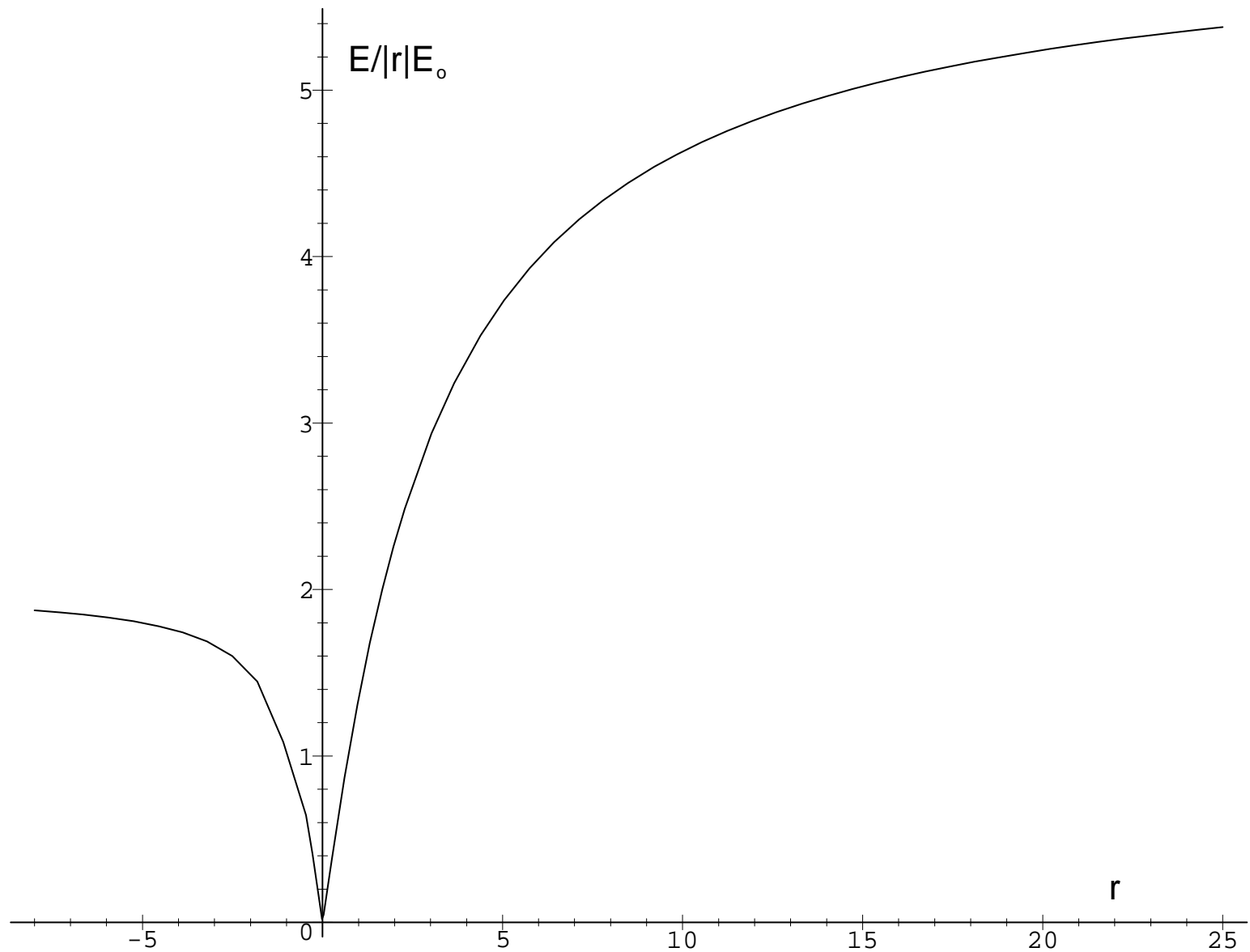
\end